\documentclass[11pt]{article}
\usepackage{jheppub}
\usepackage{amsmath,amssymb,amsfonts,graphicx,slashed,amsthm,mathtools,upgreek, enumerate, tensor, subfig}
\usepackage[dvipsnames]{xcolor}
\usepackage{arydshln}

\usepackage{comment}
\usepackage{hyperref}
\usepackage[utf8]{inputenc}
\usepackage[titletoc]{appendix}

\numberwithin{equation}{section} 
\allowdisplaybreaks

\allowdisplaybreaks

\newcommand{\be}{\begin{equation}}
\newcommand{\ee}{\end{equation}}
\newcommand{\f}{\frac}

\newcommand{\p}{\partial}

\newcommand{\bea}{\begin{eqnarray}}
\newcommand{\eea}{\end{eqnarray}}
\newcommand{\ba}{\begin{align}}
\newcommand{\ea}{\end{align}}

\newcommand{\la}{\langle}
\newcommand{\ra}{\rangle}
\newcommand{\beq}{\begin{equation}}
\newcommand{\eeq}{\end{equation}}

\DeclareMathOperator{\tr}{tr}

\title{A comment on a fine-grained description of evaporating black holes with baby universes}

\author[a]{\! Norihiro Iizuka,}
\author[b]{\! Akihiro Miyata,}
\author[c,d]{\! Tomonori Ugajin}

\affiliation[\,a]{Department of Physics, Osaka University
Toyonaka, Osaka 560-0043, JAPAN}
\affiliation[\,b]{Institute of Physics, University of Tokyo, Komaba, \\ Meguro-ku, Tokyo 153-8902, JAPAN}
\affiliation[\,c]{Center for Gravitational Physics,
Yukawa Institute for Theoretical Physics, Kyoto University,
Kitashirakawa Oiwakecho, Sakyo-ku,
Kyoto 606-8502, JAPAN}
\affiliation[\,d]{The Hakubi Center for Advanced Research, Kyoto University,
Yoshida Ushinomiyacho, Sakyo-ku, Kyoto 606-8501, JAPAN}
\emailAdd{iizuka (at) phys.sci.osaka-u.ac.jp, \\miyata (at) hep1.c.u-tokyo.ac.jp, \\
tomonori.ugajin (at) yukawa.kyoto-u.ac.jp}

\abstract{We study a  partially fine-grained description of an evaporating  black hole  by introducing an open baby universe with a boundary. Since the Page's calculation of the entropy of Hawking radiation  involves an ensemble average over a class of states,  one can formally obtain a fine-grained state by purifying this setup. For AdS black holes with a holographic dual, this purification amounts to introducing an additional boundary ({\it{i.e.,}} baby universe) and then connecting it to the original  black hole through an Einstein-Rosen bridge. We uncover  several details of this setup. As applications, we briefly discuss how this baby universe modifies the semi-classical gravitational Gauss law as well as the gravitational dressing of operators behind the horizon.
} 

\keywords{Black Holes,Conformal Field Theory, AdS-CFT Correspondence}

\preprint{OU-HET-1119, \;UT-Komaba/21-6,\; YITP-21-133}

\begin{document}

\maketitle

\section{Introduction}

Ever since Hawking discovered that a black hole has a temperature and emits thermal radiations \cite{Hawking:1975vcx,Hawking:1976ra}, how its time-evolution is consistent with the principle of quantum mechanics is one of the greatest problems in theoretical physics.
One of the key points of recent developments in quantum gravity is the role of the Euclidean wormholes, which play a crucial role in resolving the  black hole information loss problem through their non-perturbative effects (e.g., \cite{Penington:2019kki,Almheiri:2019qdq,Marolf:2020xie}).

\begin{figure}[th]
\begin{center}
\includegraphics[scale=0.3]{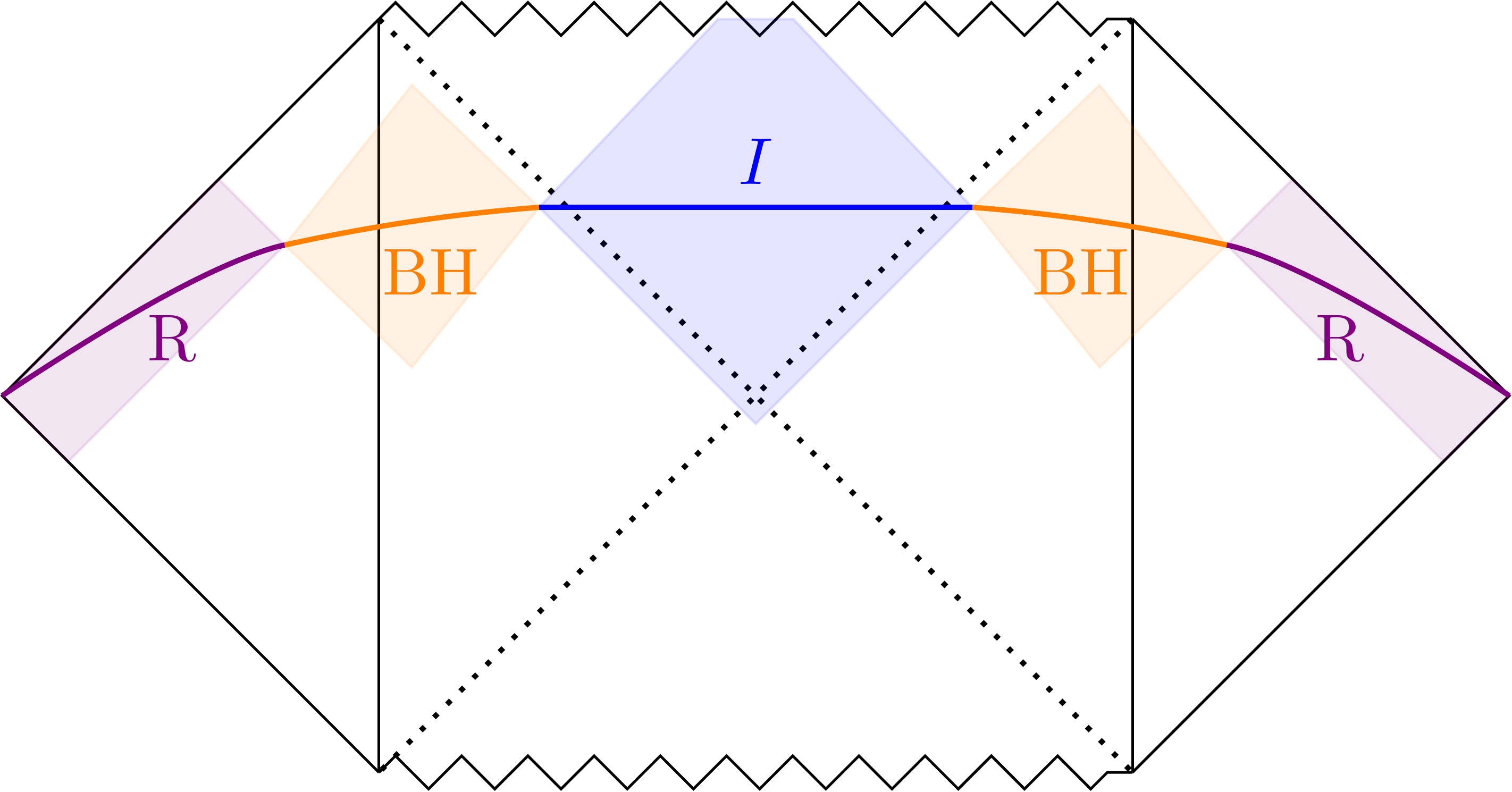}
\end{center}
\caption{The Penrose diagram of the AdS black hole attached to the non-gravitating heat bath. We take the radiation region $R$ (violet solid line) in the heat bath. After the Page time, the island region $I$ (blue solid line) becomes non-empty, and the black hole region $BH$ (orange solid line) is outside the horizon of the black hole.
The entanglement wedge of the Hawking radiation $R$ is the union of the domain of dependence of the radiation $R$ and island $I$ regions (violet and blue shaded regions), and that of the black hole $BH$ is the domain of dependence of the black hole region (orange shaded region).  
}
\label{fig:island}
\end{figure}

The von Neumann entropy of the Hawking radiation  
can be defined by the entanglement entropy of the bath region $R$, attached to the asymptotic infinity of the black hole, see figure \ref{fig:island}.  The island formula \cite{Penington:2019npb,Almheiri:2019psf,Almheiri:2019hni} tells us that this entropy is given by 
\be
S (\rho_{R}) = \underset{I}{\rm  Min Ext} \left[ \f{A(\p I) }{4G_{N}} + S_{{{\rm bulk }}} (R \cup I)\right],
\label{eq:Islandf}
\ee
where  $I$ is a region in the bulk gravitating spacetime. The  region which extremizes the above generalized entropy functional  is called the island.  This formula can be regarded as a natural extention of the RT/HRT formulae and their quantum extentions for holographic entanglement entropy in AdS/CFT \cite{Ryu:2006bv,Ryu:2006ef,Hubeny:2007xt,Faulkner:2013ana,Engelhardt:2014gca}. This island formula is indeed obtained by including so called  Euclidean replica wormholes to the gravitational path integral \cite{Penington:2019kki,Almheiri:2019qdq}. The island formula correctly reproduces the Page curve \cite{Page:1993wv,Page:2013dx} for the entanglement entropy of Hawking radiation, thus gives results consistent with the principles of quantum theory within semi-classical regime of gravity.
(See e.g., \cite{Almheiri:2020cfm,Chen:2021lnq} for reviews on this topic, and  related discussions on the island formula, e.g., \cite{Rozali:2019day,Chen:2019uhq,Bousso:2019ykv,Almheiri:2019psy,Chen:2019iro,Balasubramanian:2020hfs,Hollowood:2020cou,Alishahiha:2020qza,Chen:2020uac,Geng:2020qvw,Chandrasekaran:2020qtn,Li:2020ceg,Bousso:2020kmy,Dong:2020uxp,Hollowood:2020kvk,Chen:2020jvn,Chen:2020tes,Hartman:2020khs,Balasubramanian:2020coy,Balasubramanian:2020xqf,Ling:2020laa,Chen:2020hmv,Bhattacharya:2020uun,Harlow:2020bee,Akal:2020ujg,Hernandez:2020nem,Chen:2020ojn,Goto:2020wnk,Matsuo:2020ypv,Hsin:2020mfa,Akal:2020twv,Numasawa:2020sty,KumarBasak:2020ams,Geng:2020fxl,Deng:2020ent,Karananas:2020fwx,Wang:2021woy,Kawabata:2021hac,Fallows:2021sge,Bhattacharya:2021jrn,Kim:2021gzd,Anderson:2021vof,Miyata:2021ncm,Wang:2021mqq,Ghosh:2021axl,Aalsma:2021bit,Geng:2021iyq,Balasubramanian:2021wgd,Uhlemann:2021nhu,Qi:2021sxb,Kawabata:2021vyo,Chu:2021gdb,Langhoff:2021uct,Lu:2021gmv,Akal:2021foz,Balasubramanian:2021xcm,Ahn:2021chg,Miyaji:2021lcq,Matsuo:2021mmi,Goto:2021mbt, Anegawa:2020lzw}.)

The island formula suggests that 
the entanglement wedge of the Hawking radiation contains not only the radiation region $R$, but also the island region $I$. 
{On the other hand, the entanglement wedge of the black hole is its compliment (see figure \ref{fig:island} again).  The boundary of the island region $\partial I$ is  located just behind the  horizon for an evaporating black hole \cite{Penington:2019npb,Almheiri:2019psf,Almheiri:2019hni,Gautason:2020tmk,Hartman:2020swn}. However for an eternal black hole, it is located outside the horizon \cite{Almheiri:2019yqk,Gautason:2020tmk,Anegawa:2020ezn,Hashimoto:2020cas,Hartman:2020swn}.

In a recent interesting paper \cite{Geng:2021hlu}, such a form of entanglement wedge disconnected from the asymptotic boundary is {argued to be} inconsistent  with the long range nature of the gravitational force or equivalently diffeomorphism invariance.   Let us consider a small local  operation on the island $I$ which corresponds to a local excitation of  on the radiation Hilbert space $H_{R}$.  
In a theory of gravity, such a  small operation seems to be  detected on the asymptotic boundary of the spacetime using the gravitational Gauss law.  This is problematic because the asymptotic boundary belongs to the entanglement wedge of the black hole, {see figure \ref{fig:islandWithLocalOpe}.}
This implies that the local operation on the island region $I$ (which is supposed to be an operation on $H_{R}$) can actually change the entanglement wedge of the black hole. 

\begin{figure}[th]
\begin{center}
\includegraphics[scale=0.3]{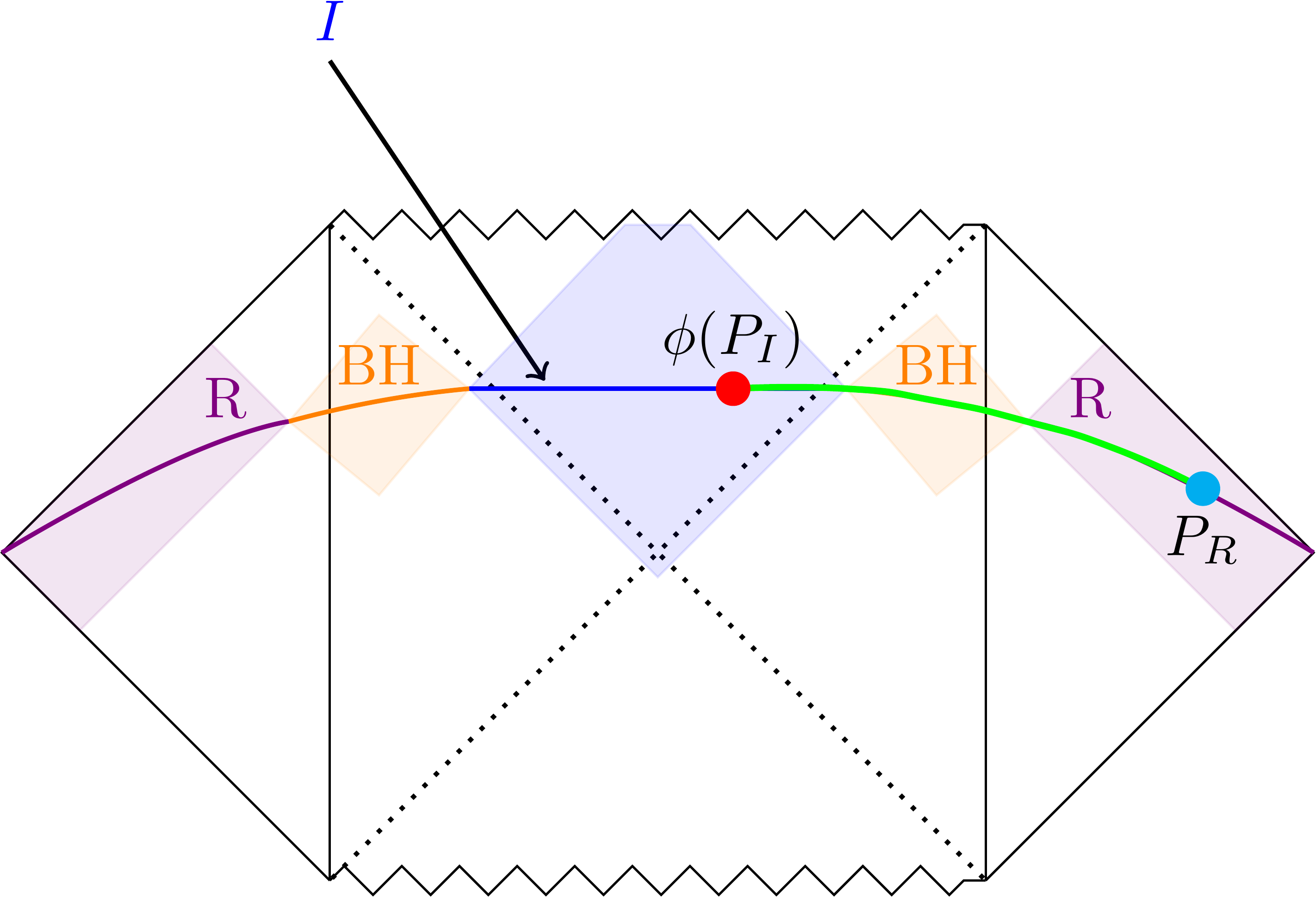}
\end{center}
\caption{The Penrose diagram of the AdS black hole attached to the non-gravitating heat bath with a small local operation $\phi(P_I)$ ($P_I \in I$, red dot) in the island region $I$.
 The entanglement wedge of the black hole $BH$ (orange shaded region) includes the asymptotic AdS boundary. The gravitational Wilson line $W(P_I,P_R)$ (green line) connecting two points $P_I$ and $P_R$ ($P_R\in R$, cyan dot)  intersects the entanglement wedge of the black hole (orange shaded region).  }
\label{fig:islandWithLocalOpe}
\end{figure}

The above apparent inconsistency of the gravitational Gauss law is related to the inconsistency of the gravitational dressing of local operators on the island region \cite{Geng:2021hlu}.  In a theory with diffeomorphism invariance, a local operator can not be physical  since it is not gauge invariant. Instead, such a local operator $\phi(P)$  needs to be ``gravitationally dressed" \cite{Donnelly:2015hta,Donnelly:2016rvo}   by attaching a Wilson line $W(P,P_B)$ connecting  the point $P$ to a point $P_B$ on the asymptotic boundary.  Then  the resulting operator $\phi(P)W(P,P_B)$ is gauge invariant. For the gravitational dressing for a local operator on the island region $P_I\in I$, which is a part of the entanglement wedge of the Hawking radiation, it is natural to choose a point on the bath region $R$ as the asymptotic boundary point $P_R\in R$. In such a case, the Wilson line $W(P_I,P_R)$ connecting the two points intersects with  the entanglement wedge of the black hole. However  the dressed operator $\phi(P_I)W(P_I,P_R)$ should belong only to the entanglement wedge of the Hawking radiation, thus this is problematic again.
{See the figure \ref{fig:islandWithLocalOpe} again.}}

In this paper, we address the above paradox by carefully examining  how the effects of random fluctuations of an evaporating black hole are geometrized in a semi-classical  description of gravity. In principle the black hole evaporation process is described by the bipartite system of  the Hilbert space of the black hole $H_{BH}$ and the one for the Hawking radiation $H_{R}$. Of course,  the description of such an entangled state involves a quantum theory of gravity, therefore it seems impossible to study such a system efficiently. However, as was first observed by Page \cite{Page:1993wv}, one can obtain a time evolution of the radiation entropy consistent with unitarity, by averaging the entropy over  the random fluctuations in the entangled state. This opens up the possibility of  having a partially fine-grained description of the evaporating black hole while maintaining its semi-classical nature,  
to the extent of getting results consistent with the principles of quantum theory. Indeed,  
 in this way, the island formula makes it possible to recover the Page curve in a semi-classical way. Specifically, the Euclidean replica wormholes nicely capture the effects of these random fluctuations and their averaging through a geometric way.

This paper concerns a  description of these random fluctuations in a Lorentzian spacetime in the semi-classical regime. We argue that the averaging over the random fluctuations can be purified by introducing an auxiliary system, often called a baby universe.  This new piece of the spacetime is connected to the original  spacetime with the black hole by an Einstein-Rosen bridge,  can be thought of as accommodating partially fine-grained information of the evaporating black hole (see figure \ref{fig:conneBH}).  See also \cite{Polchinski:1994zs, Marolf:2020rpm}  for discussions on the role of  baby universes  in the information loss paradox.

 Motivated by this observation, we then study the gravitational Gauss law in the presence of the baby universe sector. Such an introduction of the baby universe  significantly modifies the form of the gravitational Gauss law. For instance,  assuming the baby universe part has an asymptotic boundary, the gravitational Gauss law does not exactly hold within the original spacetime as there is a contribution from the baby universe sector.  
  This makes sense, because restricting our attention only  to the original black hole spacetime corresponds to  a coarse-graining. This is clearly seen from a Schwarzschild black hole solution, which has the horizon area and therefore the Bekenstein-Hawking entropy  $S_{BH}$. However it is just one solution, showing no degeneracy of the states, contradicting the huge degeneracy given by the entropy $e^{S_{BH}}$. What we expect is that in full quantum gravity, one obtains microstates of the black hole and by counting its degeneracy, one obtains $e^{S_{BH}}$. However after coarse-graining, all the details of the microstates are lost, and one cannot see the microstate differences in the Schwarzschild solution.  Assuming the dynamics of the black hole is sufficiently chaotic, two distinct energy eigenstates can never have the same energy, and the minimal value of the difference is of order $e^{-S_{BH}}$. Thus the geometric description of such a  class of microstates  by a single black hole spacetime inevitably involves a coarse-graining, in which the energy differences of order  $e^{-S_{BH}}$ are neglected.  This suggests that in the black hole spacetime, one can only trust the gravitational Gauss law  up to $O (e^{-S_{BH}})$ corrections.
  
The introduction of the baby universe with an asymptotic boundary naturally resolves the paradox of the gravitational dressing as well,  because the gravitational Wilson line starting from the island region  can now  end on the  boundary of the baby universe. This is a kind of an expected result because the island region corresponds to fine-grained information  of the evaporating black hole, so can not have a simple description within the original black hole spacetime.

The rest of the paper is organized as follows. 
In section \ref{sec:babyuni}, we study  wormholes, and the baby universe.
In section \ref{sec:modiGauss}, we present our main idea and explain how we modify the gravitational Gauss law in the presence of the baby universe. We also comment on how the boundary of the baby universe can resolve the gravitational dressing paradox. 
 In appendix \ref{sec:vNentropy}, we give the calculation of the von Neumann entropy of the Hawking radiation and that of the Hawking radiation plus the baby universe in our formalism.

\vspace{1em}
{\it Note added:} During the preparation of this paper, the papers \cite{Renner:2021qbe,Qi:2021oni,Almheiri:2021jwq} appeared, and discussed extra information coming from the ensemble nature of gravity, which is related to the baby universe degrees of freedom in our paper.

\section{Baby Universe and Ensemble nature of Semi-Classical Gravity}\label{sec:babyuni}

In this section, we clarify the role of the baby universe in the computation of the fine-grained entropy of Hawking radiation through the island formula. 

To this end, it is appropriate to begin with the fact that there are two distinct descriptions of a theory of gravity. The first one is the fine-grained description, and the second one is the coarse-grained one. 
In the first full-fledged fine-grained description of quantum gravity, we have a sufficient number of observables ({\it i.e.}, the complete set of operators of quantum gravity) to perfectly distinguish quantum states. Note that, in the description, we can perform measurements with arbitrary precision.
We are interested in the gravitational system where a black hole keeps emitting Hawking radiations. 
In a full-fledged fine-grained microscopic description, an actual state in such a system has the following form,
\begin{equation}
	|\Psi_M\rangle = \sum_{i =1}^{\mathcal{N}} \sum_{\alpha=1}^{k} {F}_{ i \alpha}^M |\psi_i \rangle_{BH} |\alpha \rangle_R \,,\label{eq:fine-microstate}
\end{equation}
where ${F}_{i \alpha}^M$ takes a {\it fixed number}.  
Here we define the orthonormal bases $|\psi_{i}\rangle_{BH}$ and $|\alpha\rangle_{R}$  of the Hilbert space $H_{BH}$ for microstates for the black hole and the similar Hilbert space $H_{R}$ for the Hawking quanta participating in the entanglement. $\mathcal{N}$ and $k$ are their dimensions.

The second description of the system is the coarse-grained one in terms of a semi-classical theory, where we have a restricted number of observables, {\it{i.e.}}, a subset of the complete set of observables of quantum gravity, or coarse-grained observables  like thermodynamical quantities. The spatial and time resolution of such observables is much larger than the Planck scale.  In this description, even by measuring coarse-grained observables precisely, we cannot completely distinguish the underlying full quantum states of the full theory, but at best a set of states with the same expectation values of the coarse-grained observables and the same semi-classical geometries.

Owing to the restricted number of observables and also to the fact that the resolution is much larger than the Planck scale, one is forced to describe the system in a coarse-grained way, in terms of 
 a {\it mixed} state, {\it i.e.}, an ensemble of states $\{p_{M}, |\Psi_M\rangle\}_{M}$\footnote{See \cite{Balasubramanian:2020lux} for a similar discussion.}. 
This ensemble consists of the class of the states $|\Psi_M\rangle$ with the {\it random coefficient} matrix $C_{ i \alpha}^M$
\begin{equation}
	|\Psi_M\rangle = \sum_{i =1}^{\mathcal{N}} \sum_{\alpha=1}^{k} C_{ i \alpha}^M |\psi_i \rangle_{BH} |\alpha \rangle_R \,.\label{eq:microstate}
\end{equation}

From the semi-classical gravity point of view, two  such states  $|\Psi_{M} \ra, |\Psi_{N} \ra$ with different random  coefficients  $C^{M}, C^{N}$ can not be distinguished. This corresponds that a coarse-grained observer describes the state in terms of the following mixed state,
\begin{equation}
	\rho_{BH\cup R}=\sum_{M} p_M |\Psi_M\rangle \langle \Psi_M| \,, \label{eq:mixedstate}
\end{equation}
where $p_M$ is the Gaussian  probability distribution determined by the ensemble of states or 
random coefficient matrix $C_{ i \alpha}^M$ as 
\begin{equation}
	\begin{aligned}
	p_M&= \left(\frac{\mathcal{N}k }{\pi}\right)^{\mathcal{N}k} \exp\left( - \mathcal{N}k \tr (C^M C^{M\dagger}) \right),
	\end{aligned}
\end{equation}
and satisfies $\sum_M p_M=1$. See \eqref{eq:PureToMix}-\eqref{eq:normal} in appendix \ref{sec:vNentropy}.  We also note that the coefficients $C_{i\alpha}$ are satisfying the following relationship,   
\begin{equation}
\begin{aligned}
        \la 1 \ra&=1\\
        \la C_{i \alpha} C^{\dagger}_{\beta j} \ra&=\frac{1}{k\mathcal{N}}\delta_{ij}\delta_{\alpha\beta}\\
        \la C_{i \alpha} C^{\dagger}_{\beta j} C_{k \gamma} C^{\dagger}_{\delta l} \ra&=\frac{1}{(k\mathcal{N})^2}\left( \delta_{ij}\delta_{\alpha\beta}\cdot \delta_{kl}\delta_{\gamma\delta}+ \delta_{il}\delta_{\alpha\delta}\cdot \delta_{jk}\delta_{\beta\gamma} \right)\\
        \la (\Pi_{a=1}^n C_{i_a \alpha_a})(\Pi_{b=1}^n C^{\dagger}_{\beta_b j_b }) \ra&=\frac{1}{(k\mathcal{N})^n}(\text{all possible contractions of indices})\\
        \la (\Pi_{a=1}^n C_{i_a \alpha_a})(\Pi_{b=1}^m C^{\dagger}_{\beta_b j_b }) \ra&=0 \quad \text{for }m\neq n
        \label{eq:haarRan}
\end{aligned}
\end{equation}
where $\la \cdot \ra$ means the average over the random coefficient matrix $C_{ i \alpha}^M$. 
The randomness  of the coefficient in  \eqref{eq:microstate}  is due to the fact that the dynamics of a black hole is highly chaotic. These can be understood as follows: 
Suppose that an observer tries to experimentally specify the fine-grained state \eqref{eq:fine-microstate}. Then the observer needs to  perform a measurement with the Planck scale precision. However, for coarse-grained observers, the resolution of the measurement is much larger than the Planck scale. Note that during the measurement time-scale, the microscopic state can evolve. 
Therefore, if the measurement time-scale is much larger than the Planck scale, the microscopic state can evolve to almost all states of the form \eqref{eq:microstate}. In this way, 
coarse-grained observers see the black hole state as \eqref{eq:microstate}. This provides an intuitive way to understand  the reason why the random matrices appear in the semi-classical description of the black hole dynamics.

Once we coarse-grain the system, the state is reduced from the pure state \eqref{eq:fine-microstate} to the mixed state \eqref{eq:microstate}, and  apparently we lose the microscopic details of the states. However we nevertheless can compute some aspects of the  {\it fine-grained} entropy  of Hawking radiation by purifying this mixed state by introducing an auxiliary system $H_{BU}$, which we often call the baby universe.  
For instance, recent progress in understanding the island formula suggests that the purification enables us to capture some part of the fine-grained  information of Hawking radiation while maintaining the semi-classical description. Discussions on  the relevance of random fluctuations for the physics of  black holes can be found  for example in  \cite{Verlinde:2012cy,Langhoff:2020jqa,Nomura:2020ska}. We also note that Gaussian random fluctuations  have a geometric interpretation in terms of end of the world branes  in two-dimensional JT gravity \cite{Penington:2019kki}.

 Note that to purify the original system with the mixed state \eqref{eq:mixedstate}, we need an auxiliary system $H_{BU}$ whose dimension is at least equal to or greater than that of the original system. The dimension of the baby universe Hilbert space depends in particular on the coarse-graining procedure.  On this new Hilbert space, the  simplest purified state is given by 
 \begin{equation}
 	| \Phi \rangle_{BH \cup R \cup BU}=\sum_M \sqrt{p_M}|\Psi_M\rangle_{BH\cup R} |M\rangle_{BU}, \label{eq:puriSta}
 \end{equation}
 where $\{|M\rangle_{BU}\}$ are orthonormal baby universe states. A fine-grained observer can access this auxiliary system, but coarse-grain observers can not. 
Let us emphasize that the description using the auxiliary system is not a full fledged fine-grained description of the system.   This is because we are artificially adding degrees of freedom, which do not show up in the original  Hilbert space $H_{BH} \otimes H_{R}$. More concretely, in the quantum gravity description,  the actual fine-grained state realized in the system is one of the states in the ensemble, not the one  with the baby universe. We nevertheless consider the purified state \eqref{eq:puriSta}, because it has an effective semi-classical description, on the contrary to the full fledged fine-grained state in quantum gravity. Furthermore, as we will show later, if we are only concerned with the averaged property of the fine-grained entropy, such as  the Page curve,  considering this  purified state is good enough.

Note that by tracing out the black hole degrees of freedom $BH$ in the mixed state \eqref{eq:mixedstate}, the reduced density matrix of the Hawking radiation $\rho_R$ gives an approximately thermal mixed state, and the von Neumann entropy $S_{\rm{vN}}[\rho_R]$ gives the Hawking's result 
\begin{equation}
\begin{aligned}
   S_{\rm{vN}}[\rho_R]&= S_\mathrm{vN}[\; \la \rho_{(M)R} \ra_{M} \; ]\\
   &=\log k,
 \end{aligned}\label{eq:vNcoase}
\end{equation}
where  we have defined, 
\be
\rho_{(M)R}=\tr_{BH}\left[ \, |\Psi_M\rangle \langle \Psi_M|_{BH\cup R} \, \right],  \quad \la \rho_{(M)R} \ra_{M}=\sum_{M} p_{M} \; \rho_{(M) R}. 
\ee
See the appendix \ref{subsec:Hawki} for detailed derivation.

 Now let us consider the same  entropy of Hawking radiation in the fine-grained description. To do so, let us first figure out a geometric description of the purified state
 \eqref{eq:puriSta}. In this state, the Hawking radiation $H_{R}$ and the black hole $H_{BH}$ are entangled with the auxiliary baby universe $H_{BU}$. 
From the viewpoint of ER=EPR \cite{Maldacena:2013xja}, we expect that this is realized geometrically by an Einstein-Rosen bridge connecting the baby universe and  the original system (see figure \ref{fig:conneBH}). The property of the ER bridge  depends highly on the choice of the ensemble.  If we realize this system within the framework of the AdS/CFT correspondence, the auxiliary universe can be modeled by an additional boundary and its gravity dual involves an Einstein-Rosen bridge connecting the new boundary\footnote{See \cite{Heckman:2021vzx} for a similar discussion by using sting theory.}.  This purification process is the key in recent studies, especially in the finding of the island formula  which captures {some aspects} of fine-grained information of the quantum gravity states, in the semi-classical description through a non-perturbative way. For instance, in describing an evaporation process of a black hole  semi-classically, such non-perturbative contributions are required to get a consistent result. In such a process  discreteness of the energy spectrum of the black hole microstates is a crucial ingredient to ensure  unitarity of the process. However, in the coarse-grained description,
energy differences between black hole micro-states  are invisible, since they are typically of order $\mathcal{O}(e^{-S_{BH}})$, where 
$S_{BH}$ is the Bekenstein-Hawking entropy \cite{Balasubramanian:2006iw}. 
 A discrete energy spectrum is only  after including 
non-perturbatively small contributions which are provided by Euclidean wormholes \cite{Maldacena:2001kr,Saad:2019lba}.

What the island formula implies is that 
 one should identify the fine-grained Hilbert space of the Hawking radiation  $H_\mathbf{R}$  with the tensor product of two Hilbert spaces $H_{R}\otimes H_{BU}$, after the Page time. On the other hand, before the Page time $H_\mathbf{R}$ should be identified with just that of the Hawking radiation $H_{R}$, and correspondingly the Hilbert space of the black hole should coincide with the tensor product of the black hole and the baby universe $H_{BH}\otimes H_{BU}$. 
 This difference between the radiation Hilbert spaces before and after the Page time comes from the fact that the inequality for the dimensions of the Hilbert spaces of   the Hawking radiation and that of the black hole changes. In fact,  before the Page time, since the total state \eqref{eq:puriSta} is pure, the von Neumann entropy of the union of the black hole and the baby universe $BH \cup BU$ is equal to the previous von Neumann entropy \eqref{eq:vNcoase} of $R$, {\it i.e.},  $S_\mathrm{vN}[\rho_{BH\cup BU}]=S_\mathrm{vN}[\rho_{R}]=\log k$, which is consistent with the island formula before the Page time.

After the Page time, the reduced density matrix of the Hawking radiation and the baby universe $\rho_{R\cup BU}$  in \eqref{eq:puriSta}  gives the the fined grained entropy of the Hawking radiation, which deviates from the entropy \eqref{eq:vNcoase} of the naive density matrix \eqref{eq:mixedstate},
 \begin{equation}
	\begin{aligned}
		S_\mathrm{vN}[\rho_\mathbf{R}]&=S_\mathrm{vN}[\rho_{R\cup BU}]\\
		&=\log \mathcal{N}\\
		&=S_{BH}. 
	\end{aligned}
	\label{eq:vNfine}
\end{equation}
In appendix \ref{subsec:HawkiBaby} we provide  details of  this calculation. {The result reproduces the behaviour of the Page curve after the Page time, giving the Bekenstein-Hawking entropy $S_{BH}$.  Therefore by appropriately dividing the total system $BH\cup R \cup BU$, we can get the von Neumann entropy  which obeys the Page curve (see table \ref{ta:howtoDiv})\footnote{ One may also consider the possibility of 
dividing the baby universe Hilbert space $H_{BU}$ into two parts $H_{BU_{BH}}\otimes H_{BU_{R}}$, and then define the radiation Hilbert space as $H_{\bf R}=H_{BU_{R}} \otimes H_{R}$, instead of $H_{\bf R}=H_{BU} \otimes H_{R}$ which we do in the body of the paper.
In such a case, the states of the baby universe are given by $|M \ra_{BU_{BH}}\otimes |M \ra_{BU_{R}}$.
In this case, assuming the orthogonality of the basis of $H_{BU_{BH}}$,  we see that the entropy of $\rho_{BU_{R} \cup R }$ is given by 
\be
S_\mathrm{vN}[\rho_{BU_{R} \cup R }] = -\sum_{M} p_{M} \log p_{M} + \sum_{M} p_{M} S_\mathrm{vN} [\rho_{(M)R}],
\ee
where $\rho_{(M)R}$ is the reduced density matrix given by \eqref{eq:ReducedR}. Then  it is natural to define the fine grained entropy of Hawking radiation $S(\rho_{{\bf R}})$ as a conditional entropy of knowing the probability distribution  ${p_{M}}$ by subtracting the classical Shannon piece $H (p_{M}) =-\sum_{M} p_{M} \log p_{M}$
\be
S(\rho_{{\bf R}})= S_\mathrm{vN}[\rho_{BU_{R} \cup R }] -H (p_{M}) =  \sum_{M} p_{M} \; S_\mathrm{vN} [\rho_{(M)R}].
\ee
However we do not know the natural choice for such a splitting of the baby universe Hilbert space $H_{BU}$.}.

\begin{table}[h]
{
\centering
\begin{tabular}{|l|c|c|c|}
\hline
 & Black Hole & Hawking Radiation  & von Neumann Entropy \\ \hline
Before the Page time & $BH \cup BU$ & $R$ & $S_\mathrm{vN}[\rho_{BH\cup BU}]=S_\mathrm{vN}[\rho_{R}]=\log k$ \\ \hline
After the Page time & $BH $ &  $R\cup BU$ & $S_\mathrm{vN}[\rho_{BH}]=S_\mathrm{vN}[\rho_{R\cup BU}]=S_{BH}$ \\ \hline
\end{tabular}
\caption{How to divide the total system $BH\cup R \cup BU$ into two sub systems before and after the Page time, and the corresponding von Neumann entropies.}
\label{ta:howtoDiv}}
\end{table}

At the same time, we know that the fine-grained entropy $S_\mathrm{vN}[\rho_\mathbf{R}] $ of  Hawking radiation is computed by the island formula \eqref{eq:Islandf} too. In the entropy calculations using this formula, it was crucial to include the contribution of the  island, which typically occupies the region behind the horizon of the black holes. Therefore it is natural to identify the island region behind the horizon with the Einstein-Rosen bridge  of the purified state \eqref{eq:puriSta} connecting the  
original spacetime and the baby universe,  which stores fine-grained information of the original spacetime.

These states $\{ | M \ra_{BU} \}$ in the fine-grained Hilbert space can be naturally identified with so called $\alpha$ states \cite{Coleman:1988cy,Giddings:1988cx,Giddings:1988wv} in the baby universe Hilbert space which diagonalizes the boundary creation operators \cite{Marolf:2020rpm}. Then each fine-grained state $|\Psi_{M}\ra  | M \ra$ belongs to different superselection sector, because  each  $\alpha $ state does. In particular, this means that  off diagonal element of matrix  $\la \Psi_{M}| \la M| (\mathcal{O}\otimes I )|\Psi_{N}\ra  | N \ra $  for any local operator $\mathcal{O}$ on the black hole $BH$ and the Hawking radiation $R$ vanishes, therefore any local measurement on them can not distinguish the entangled pure  state \eqref{eq:puriSta} with the mixed state only with classical correlation of the following form 
\be 
\rho  = \sum_{M} p_{M}\;  | \Psi_{M} \ra \la   \Psi_{M}  | \otimes | M \ra \la M|, \label{eq:TMD}
\ee
in the sense that
\be
\tr \left[|\Phi\ra \la \Phi | \, (\mathcal{O}\otimes I ) \right] = \tr \left[\rho\, (\mathcal{O}\otimes I )\right]=\sum_M p_M \la \Psi_M | \mathcal{O} |\Psi_{M}\ra.\label{eq:thermalAve}
\ee
In other words, LOCCs acting only on the black hole $BH$ and the Hawking radiation $R$, which can be available to coarse-grained observers, can not distinguish the classically and quantum mechanically correlated states \eqref{eq:puriSta}, \eqref{eq:TMD}.
However one can easily see the entanglement entropy of  these two states on $ \mathbf{R}= R \cup BU$ are different. Indeed, the entropy of $\rho$ contains a classical Shannon term, whereas the same entropy of \eqref{eq:puriSta} does not. From another point of view, LOCCs on the sub-system $BH\cup R$ and the baby universe $BU$, which can only be available to fine-grained observers, can distinguish the classically and quantum mechanically correlated states, since the equalities in \eqref{eq:thermalAve} do not necessarily hold for operators on $BH\cup R \cup BU$.

In the next section,  we discuss several  properties of the baby universe and the wormhole connecting the baby universe and the original spacetime. 
The wormholes may be dependent on the actual geometry of  the baby universe. We cannot fully specify the geometry of  the baby universe from the first principles of quantum gravity. There is a canonical and minimal choice for such a baby universe; starting from the original system $|{\Psi}_M\rangle$, we prepare a copy of it  $|\tilde{\Psi}_M\rangle$, and regard it as a purifier  $|M\rangle_{BU} =|\tilde{\Psi}_M\rangle_{\text{Puri.}} $.
  Then the expression \eqref{eq:puriSta} becomes 
 \begin{equation}
 	\sum_M \sqrt{p_M}|\Psi_M\rangle_{BH\cup R} |\tilde{\Psi}_M\rangle_{\text{Puri.}}.\label{eq:puriCopy}
 \end{equation}
 The existence of the boundaries in the original system $|{\Psi}_M\rangle$ implies that purifier $|M\rangle_{BU} =|\tilde{\Psi}_M\rangle_{\text{Puri.}} $ should also have boundaries. 
 More generally, there is a possibility that we may choose the multiple copies of the original system as the baby universe $|M\rangle_{BU} =|\tilde{\Psi}_M\rangle_{\text{Puri.}}^{\otimes n} $ and further choose their linear combinations as that. 
	Again from ER=EPR this entanglement between the two spacetimes implies the existence of the wormhole connecting two island regions for two spacetimes. This wormhole will affect the  non-perturbative physics of this system. Note that the more the number of copies of the original spacetime increase, the more the effects from wormholes are topologically suppressed.

\section{Gauss Law modified by the Baby Universe}\label{sec:modiGauss}

In this section, we discuss the physical consequences of the existence of the baby universe sector introduced in the last section, which accommodates fine-grained information of the system. We are mainly interested in  how the baby universe helps to recover information of the black hole interior from Hawking radiation. We will also briefly mention the relation between our discussion and the paradox raised in the recent paper \cite{Geng:2021hlu}.  

Before doing so, let us present a  remark. In the light of AdS/CFT correspondence, the introduction of an  additional boundary, {\it i.e.}, the boundary of the baby universe sounds puzzling, because AdS/CFT is the correspondence between a theory of  full quantum gravity in the bulk and a (non gravitating) CFT
on the boundary. This means that in principle, all the details of the bulk quantum gravity Hilbert space can be read off from the single  CFT Hilbert space.
Therefore, we do not need the second copy of the CFT, as we did in the previous section, which results in the baby universe sector.

Nevertheless, we are forced to do so, because we are sticking to a semi-classical description of the system. Then, to restore fine-grained information within the semi-classical regime, we need to introduce an auxiliary system  and regard the new degrees of freedom as a part of the radiation degrees of freedom after the Page time. If we do not do this, this restriction amounts to that on the boundary, we are only accessible to a sub-Hilbert space $H_{{\rm coarse}}$ which  characterizes coarse-grained degrees of freedom. To incorporate the rest of the CFT Hilbert space, which we term $H_{{\rm fine}}$ just because it describes fine-grained degrees of freedom, we need to introduce a second copy of the CFT Hilbert space, and accommodate $H_{{\rm fine}}$  to it.

The full Hilbert space on the single boundary is obtained by gluing two asymptotic boundaries of the spacetime. 
In the resulting bulk spacetime, there are two homologically  inequivalent paths, both of which connect  a point in the interior of the black hole (and belong to the island region)
to the boundary of the spacetime (see figure \ref{fig:identifiedBH}). The first path is the trivial one (the blue line in figure \ref{fig:identifiedBH} ), which entirely lies within the original spacetime. This path necessarily intersects with the entanglement wedge of the black hole. However, in the presence of the baby universe, there is a second path which does not cross the entanglement wedge of the black hole. Instead, it crosses the
 Einstein-Rosen bridge connecting the original spacetime to the baby universe, and reaches the second asymptotic boundary which accommodates fine-grained degrees of freedom as in the green line in figure \ref{fig:identifiedBH}. Since these two boundaries are in the end glued together, it connects the island region and the conformal boundary, without passing through the entanglement wedge of the black hole.

\begin{figure}[th]
\centering
\includegraphics[scale=0.28]{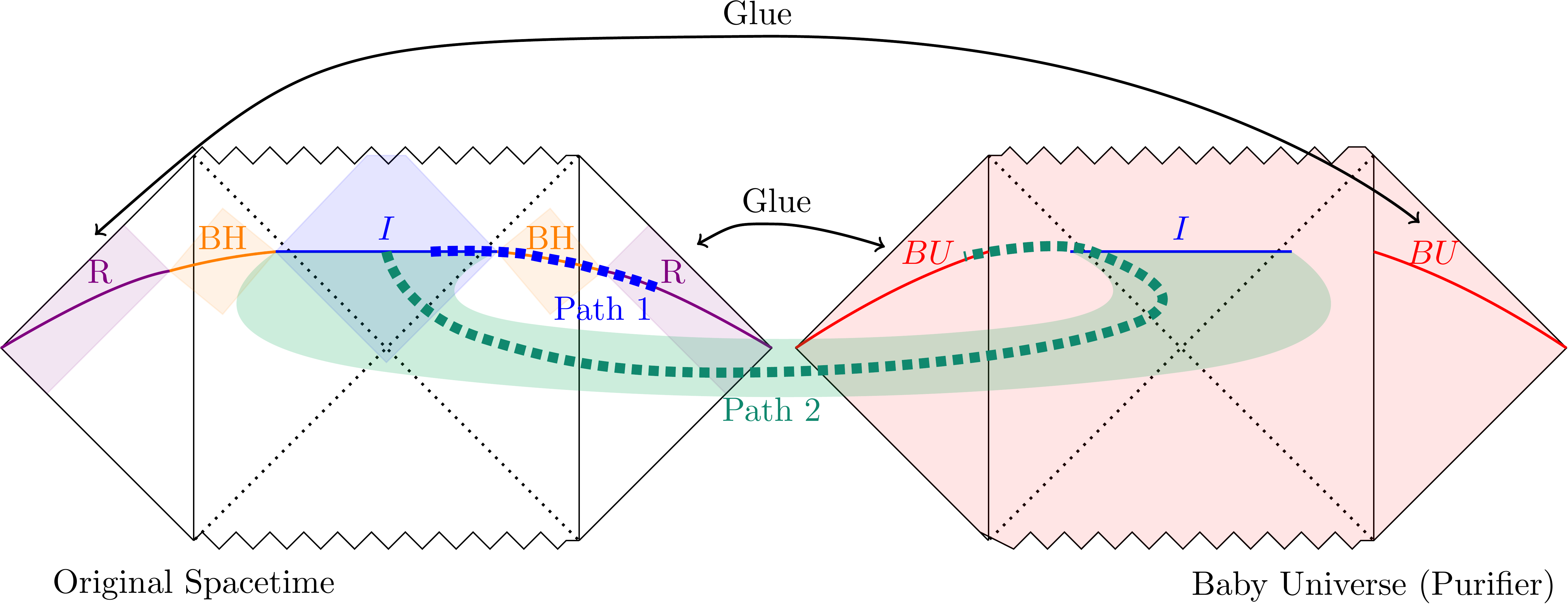}
\caption{Schematic picture of the geometry of the AdS black hole coupled the bath CFT (left Penrose diagram) and the baby universe geometry (right red Penrose diagram) connected by the Einstein-Rosen bridge (transparent green shaded region), corresponding to the state (\ref{eq:puriCopy}). After the Page time, the fine-grained Hawking radiation $\mathbf{R}$ is the union of the Hawking radiation $R$ (violet region) and the baby universe $BU$ (red region). We regard the above spacetime describing this union  by gluing two distinct asymptotic boundary regions $BU$ and $R$ . The island region $I$ is connected to the fine-grained Hawking radiation $R\cup BU$ through two paths, path 1 and 2. The path 1 (thick blue dotted line) intersects with the entanglement wedge of the black hole $BH$ (orange shaded region), but the path 2 (thick green dotted line) does not intersect with that.}
\label{fig:identifiedBH}
\end{figure}

\subsection{The modification of Gauss Law}

In the presence of the baby universe sector which has its own asymptotic boundary, the 
gravitational Gauss law is inevitably modified. Let $\Sigma$ be a time slice of the spacetime, then the gravitational Gauss law  relates the expectation value of the bulk stress energy tensor   $\la T_{bulk} \ra$  to the boundary energy $H_{\p}[h]$ (holographic stress energy tensor) 
\be
    \la T_{bulk} \ra = H_{\p}[h].\label{eq:gaussorig}
\ee
Here the boundary energy $ H_{\p}[h]$ is explicitly given by the integration of the ADM current $J^{i}$ over the conformal boundary $\p \Sigma$ \cite{Chowdhury:2021nxw},
\begin{equation}
    \begin{aligned}
        H_{\p}[h]&\equiv \frac{1}{2 \kappa^2 } \int_{\p \Sigma} d^{d-1} x\sqrt{g}\; n_i J^{i} \qquad  (\kappa =\sqrt{8 \pi G_{N}}),
    \end{aligned}
\end{equation}
where $n_i$ is the normal vector to the boundary $\p \Sigma$, and the ADM current $J^{i}$ is defined by
\begin{equation}
J_{i} \equiv N \nabla^{j}\left(h_{i j}-h g^{0}_{i j}\right)-\nabla^{j} N\left(h_{i j}-h g^{0}_{i j}\right)
\end{equation} 
under the ADM decomposition 
\begin{equation}
d s^{2}=-N^{2} d t^{2}+g_{i j}\left(d x^{i}+N^{i} d t\right)\left(d x^{j}+N^{j} d t\right),
\end{equation}
and the expansion from the background metric $g_{i j} = g^{0}_{i j} + \kappa h_{i j}$.
More precisely, \eqref{eq:gaussorig} is a perturbative version of the gravitational Gauss law which can be derived  from the full Hamiltonian constraint 
\be
\begin{aligned}
    \mathcal{H}[\pi_{ij}, g_{ij}]=2 \kappa^{2} g^{-1}\left(g_{i j} g_{k l} \pi^{i k} \pi^{j l} -\frac{1}{d-1}\left(g_{i j} \pi^{i j}\right)^{2}\right)-\frac{1}{2 \kappa^{2}}(R-2 \Lambda)+\mathcal{H}^{\text {matter}}=0, 
\end{aligned} \label{eq:constraint} 
\ee
where $g_{ij}$ is the metric on the Cauchy slice, $\pi_{ij}$ is the conjugate momentum, and $\mathcal{H}^{\text {matter}}$ is the matter Hamiltonian density. 
Expanding \eqref{eq:constraint} from the background  metric, $g_{i j} = g^{0}_{i j} + \kappa h_{i j}$, then look at the second order of the expansion gives \eqref{eq:gaussorig}.
Details of the derivation can be found, for example in \cite{Chowdhury:2021nxw}. $H_{\p}[h]$ should be understood as the change of the  mass of the black hole, $H_{\p}[h] =M_{BH}[g+h]-M_{BH}[g]$ due to the back reaction from the bulk stress energy tensor, $\la T_{bulk} \ra$.
      
 In the paper \cite{Geng:2021hlu}, it was argued that the gravitational Gauss law  provides an interesting puzzle on the island formula.   Suppose we act  a local operation on a state on  the island region. Since the information of the island region is encoded in the Hilbert space of Hawking radiation $H_{R}$, this operation can be regarded as a local operation on $H_{R}$. This operation changes the expectation value of the bulk stress energy tensor. Then the gravitational Gauss law  relates this change  of  $\la T_{ bulk} \ra$ on the island region behind the horizon to the change of  the boundary  energy $H_{\p}$.  This means that  any change on the island region,  no matter how it is small,  is always detectable from the conformal boundary $\p \Sigma$. However, this sounds troublesome because  $\p \Sigma$  belongs to the entanglement wedge of the black hole. For instance, this implies that  in the  bipartite system $H_{R} \otimes H_{BH}$,  a local operation on $H_{R}$  can change the state of $ H_{BH} $.
   
 The above paradox is naturally resolved, once we take into account the effects of the baby universe sector which admits the new boundary (see figure \ref{fig:identifiedBH}).   In the presence of this new part of the spacetime, the gravitational Gauss law must be modified as 
   \be
   \la T_{bulk} \ra = H_{\p BH}[h] + H_{\p BU}[h],
   \label{eq:gengauss}
   \ee
where  we denote      
$H_{\p BH}[h]$ by the boundary energy of the original spacetime with the black hole, and similarly $H_{\p BU}[h]$ is the boundary energy of the baby universe.

This form of the gravitational Gauss law immediately implies that, in the presence of the baby universe,  operations on the island region need {\it not} to be detected on the conformal boundary of the black hole. In other words,  $\la T_{bulk} \ra \neq 0$ does not necessarily imply $H_{\p BH}[h] \neq 0$. Rather, it is natural to  relate  $\la T_{bulk} \ra $ on the island region to the boundary energy of the baby universe $H_{\p BU}[h]$ because the island region is encoded to the Hilbert space of fine-grained Hawking radiation $H_{\mathbf{R}}= H_{R} \otimes H_{BU}$. 
Indeed, the island region encodes fine-grained information of Hawking radiation after the Page  time, so from the boundary point of view such bulk operations on this region should be encoded in the  fine-grained part of the CFT Hilbert space, which coincides with the boundary Hilbert space of the baby universe.

Another way to put this is the following. Let us consider putting a local operator in the spacetime. The gravitational Gauss law implies that by measuring the {\it total} flux for an appropriate closed surface we can know the ``mass'' of the particle within the closed surface. The non-perturbative gravitational effect from the wormhole makes the measurement of the flux highly non-trivial. The wormhole can release some part of the flux of the original spacetime into the purifier (see figure \ref{fig:schematicPic}). Here we note that since in our setup the baby universe has boundaries, flux lines can end on the boundaries of the baby universe as figure \ref{fig:schematicPic}. Namely, in measuring the {\it total} flux, we also need to consider the purifier (right spacetime of figure \ref{fig:schematicPic}) or equivalently the baby universe in addition to the original spacetime (left spacetime of figure \ref{fig:schematicPic}). By the usual gravitational Gauss law, if we just measure the flux of the original spacetime only (left spacetime of figure \ref{fig:schematicPic}), then we cannot specify the exact mass.  
The modification is not visible within the coarse-grained precision. However, without the modification, we may encounter many problems, e.g., violation of the conservation law.

There are  several other implications of the generalized Gauss law  \eqref{eq:gengauss} as well. First, 
the existence of the baby universe boundary energy 
term indicates that the gravitational  Gauss law does not precisely  hold within the original black hole spacetime, $\la T_{bulk} \ra \neq  H_{\p BH}[h] $ in general. 
For instance, one way to think about the generalized Gauss law of the form  \eqref{eq:gengauss} is, it relates the spectrum of the  fine-grained part $H_{\p BU}[h]$ to the coarse-grained part 
$H_{\p BH}[h]$. We expect that the fine-grained part is discrete, and the typical differences between two nearest energy eigenvalues are of order $e^{-S_{BH}}$.  This forces the coarse-grained part 
also discrete, which is necessary for  unitary time evolution. 

Let us estimate the magnitude of the violation of the gravitational Gauss law in the black hole spacetime. 
 In order to obtain a unitary time evolution  of an evaporating black hole,   we need   non-perturbative effects of order  $e^{-S_{BH}}$, where $S_{BH}$ is the entropy of the black hole.  This means that  we need fine-grained states in  a small energy  window of order $e^{-S_{BH}}$ , thus $H_{\p BU}$ is  of the same order. This leads us to the conclusion that 
\be
\la T_{bulk} \ra - H_{\p BH}[h] =O(e^{-S_{BH}}),
\ee
{\it i.e.}, the gravitational Gauss law is violated only non-perturbatively.

\begin{figure}[th]
\includegraphics[scale=0.27]{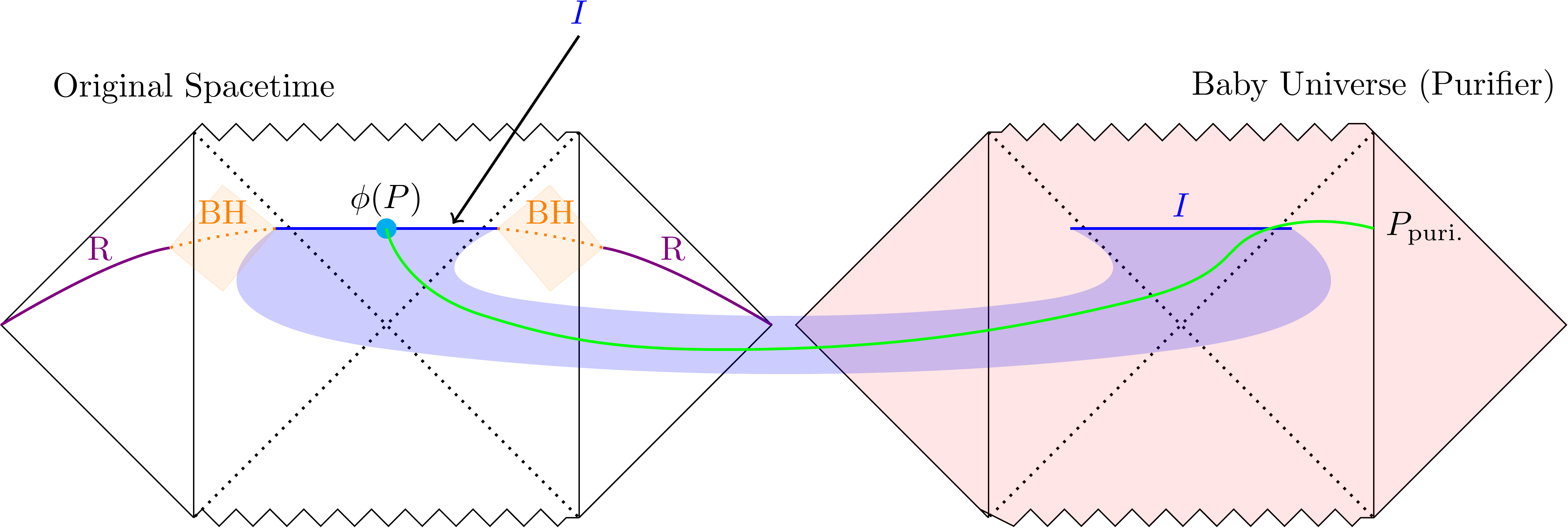}
\caption{Schematic picture of the geometry of the AdS black hole coupled the bath CFT (left Penrose diagram) and their copy (right Penrose diagram) connected to the original spacetime through the wormhole (blue region), corresponding to the state \eqref{eq:puriCopy}. The local operator $\phi$ in the island (cyan dot) can be gravitationally dressed with a gravitational Wilson line $W_\text{gravity}(P,P_\text{puri.})$ (green line) which ends on the baby universe (right Penrose diagram) without intersecting the entanglement wedge of the original black hole degrees of freedom (orange shaded region). 
}
\label{fig:conneBH}
\end{figure}

\begin{figure}[th]
\centering
\includegraphics[scale=0.27]{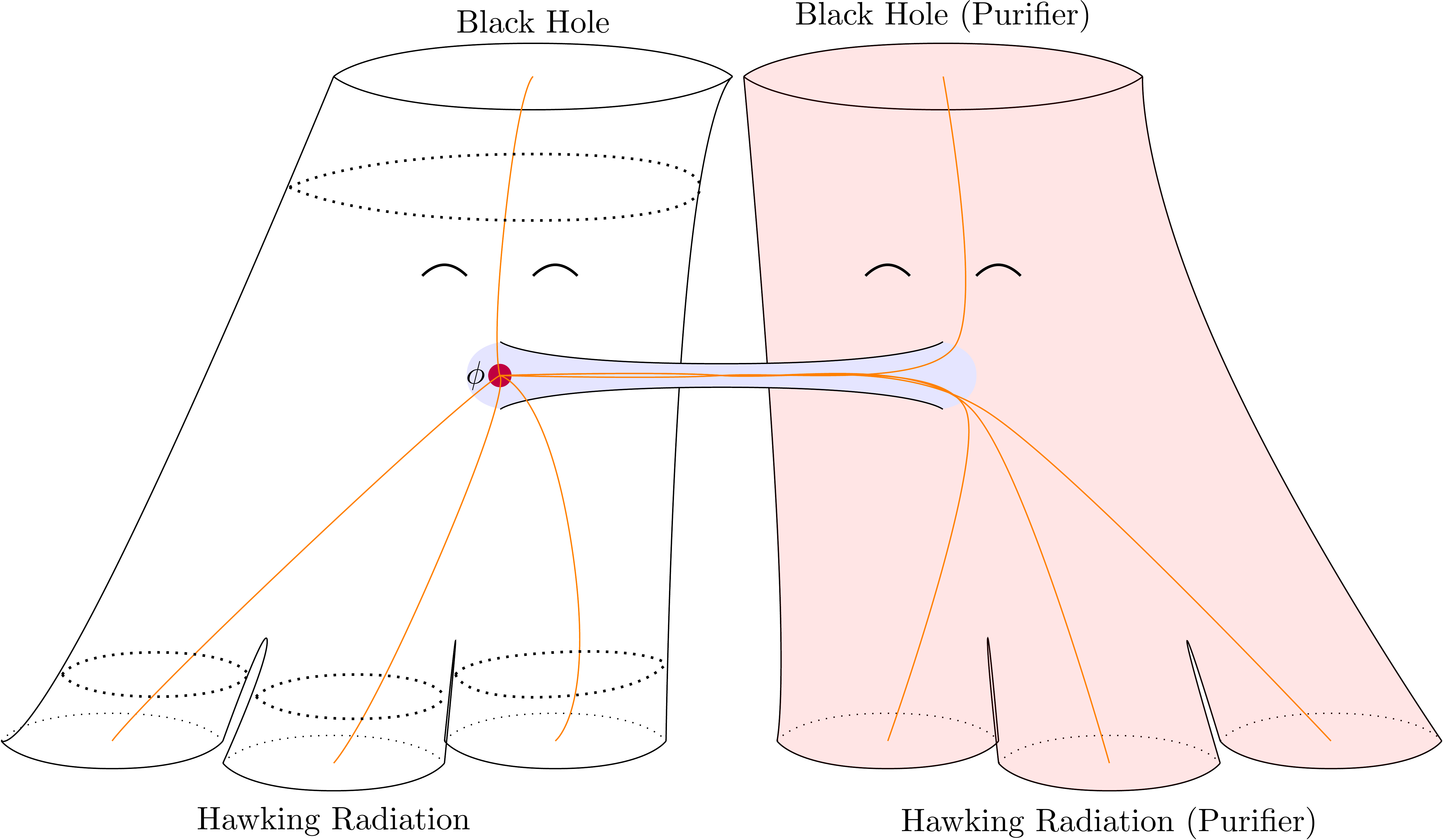}
\caption{Schematic picture of flux lines on the geometry corresponding to (\ref{eq:puriCopy}) (, similar to the figure discussed in \cite{Akers:2019nfi}). The dotted lines are horizons. A local operator $\phi$ (red dot) is put at the original spacetime (left region).  The two spacetimes are connected by the Einstein-Rosen bridge  (blue region). Some of flux lines (orange lines) escape into the other spacetime (right red region) through the Einstein-Rosen bridge .}
\label{fig:schematicPic}
\end{figure}

 We should emphasize that such a baby universe is different from those appearing by cutting Euclidean wormholes into half, in the semi-classical gravitational path integral.  Such a baby universe 
  is always closed and does not have any asymptotic boundary.  Such a closed  universe, corresponds to an additional factor of the von Neumann algebra of the CFT \cite{Gesteau:2020wrk}.  On the contrary to this, our baby universe has an asymptotic boundary to encode the  fine-grained information of the state. It would be interesting 
  to further investigate the relation between the two.

  We also speculate the realization of fine-grained degrees of freedom in terms of a baby universe with a boundary has an interesting application to the physic of a closed universe. Sometimes it is argued that the Hilbert space of such a closed universe is one-dimensional  \cite{McNamara:2020uza,Gesteau:2020wrk}\footnote{Sometimes this problem in $d$($\geq4$)-dimensional spacetime is called the baby universe hypothesis \cite{McNamara:2020uza}.}, because in  the absence of boundary, the left hand of  the gravitational Gauss law \eqref{eq:gaussorig} is always zero, so any operations are not allowed at all. However, as we saw above, one way to obtain  its fine-grained description is to connect it to an open baby universe with a boundary. Then 
  the generalized Gauss law \eqref{eq:gengauss} does  allow operations on the baby universe boundary only. It would be interesting to explore further implications of the observation.

 \subsection{Comment on gravitational dressing}

In a theory with dynamical gravity,  a local operator is not physical, since it is not diffeomorphism invariant. One way to make it diffeomorphism invariant is  to connect the local point $P$ to a point $P_{\p}$  on the boundary, via a  gravitational Wilson line, {\it i.e.,}  $\phi(P) \rightarrow \phi(P) W_\text{gravity}(P,P_{\p})$. This prescription is called gravitational dressing. In \cite{Geng:2021hlu} it was argued that such a gravitational dressing of a local operator on the island region leads to an inconsistency of the island prescription. This is because the relevant gravitational Wilson line connects a point on the island  to a point on the conformal boundary of the AdS black hole. However, this sounds puzzling, because  whereas the island prescription asserts an operator on the island region locally acts on the radiation Hilbert space, the gravitational Wilson line attached to it  enters the entanglement wedge of the black hole, thus it does change the state of $H_{BH}$. 

In our point of view, the above paradox is naturally resolved, since in the presence of the baby universe with a boundary, the gravitational Wilson line can end on this (see figure \ref{fig:conneBH}).  Furthermore  since this new boundary belongs to the radiation degrees of freedom after the Page time, it is still an operator on the radiation Hilbert space, even after the gravitational dressing.

\section{Discussion}\label{sec:discussion}

In this paper, we studied a partially fine-grained description of semi-classical  
evaporating black holes, by introducing an auxiliary system called a baby universe. We argued  that in usual consistent long-range gravitational theories, the gravitational Gauss law must be modified by the baby universe connected to the original spacetime, and when there is an island, this   modification is crucial to get results consistent with the idea of the entanglement wedge reconstruction.

It would be interesting to study concrete geometric models, to understand further detailed properties of this system. A class of  the candidate geometries is the multi-boundary wormhole solution of three-dimensional Einstein gravity with a negative cosmological constant. It is convenient to use the coordinates in which the metric of $AdS_{3}$ takes the following form
\be
ds^{2} =-dt^{2} +\cos^{2} t d\Sigma_{2}^{2},
\ee
where $d\Sigma_{2}^{2}$ denotes the metric of two-dimensional hyperbolic space. These multi boundary wormholes are constructed by taking appropriate quotient of the hyperbolic space by the isometry group $SL(2,R) \times SL(2,R)$.  
Such a geometry has multiple conformal boundaries, on each of which  we can define a CFT Hilbert space. In each asymptotic region, there is a horizon, whose area counts the number of degrees of freedom  in the boundary Hilbert space. For simplicity, below let's consider such a geometry with three asymptotic boundaries. These three boundaries represent   the Hilbert space of Hawking radiation $H_{R}$, the black hole $H_{{\rm BH}}$, and the baby universe $H_{{\rm BU}}$.   Thus, one can identify the horizon area of each asymptotic region with the entanglement entropy of each Hilbert space computed in  \eqref{eq:SEERadiation},   \eqref{eq:SEEBH}, \eqref{eq:SEEBU} in Appendix \ref{sec:vNentropy}. The region behind these horizons is identified with the Einstein-Rosen bridge which connects the original black hole with the baby universe,  discussed in the body of this paper. The geometric description manifests the following entanglement structure of \eqref{eq:puriSta}. When $k ={\rm dim} H_{R}$  is small, which models the beginning of the black hole evaporation,  this system is almost a bipartite in which $H_{{\rm BH}}$ and $H_{{\rm BU}}$ are entangled. As we increase $k$, the cross section of the ER bridge gets larger, and  at sufficiently late times  $k \gg 1$, $H_{{\rm BU}}$ becomes  mostly entangled with the radiation Hilbert space $H_{{\rm R}}$.  
This means that the state  \eqref{eq:puriSta}  is reconstructable  from the two Hilbert spaces $H_{R}$ and $H_{{\rm BU}}$.

\section*{Acknowledgement}

TU thanks Kanato Goto,Yuka Kusuki, Yasunori Nomura,  Kotaro Tamaoka and Zixia Wei for useful discussions in the related projects. 
The work of NI was supported in part by JSPS KAKENHI Grant Number 18K03619. 
TU was supported by JSPS Grant-in-Aid for Young Scientists 19K14716.  
NI and TU were also supported by MEXT KAKENHI Grant-in-Aid for Transformative Research Areas A “Extreme Universe” No.21H05184.

\appendix

\section{The von Neumann Entropy of the naive Hawking Radiation, the Black hole and the Baby Universe}\label{sec:vNentropy}

In this appendix, we give von Neumann entropies of various subsystems for the states \eqref{eq:puriSta}, \eqref{eq:mixedstate} by using the relationship \eqref{eq:haarRan}. In particular, we calculate the von Neumann entropies of three cases: (i) the naive Hawking radiation or the union of the black hole and the baby universe, $S_\mathrm{vN}[\rho_R]=S_\mathrm{vN}[\rho_{BH\cup BU}]$; (ii) the naive Hawking radiation and the baby universe or the black hole $S_\mathrm{vN}[\rho_{R\cup BU}]=S_\mathrm{vN}[\rho_{BH}]$; (iii) the baby universe or the union of the black hole and the naive Hawking radiation $S_\mathrm{vN}[\rho_{BU}]=S_\mathrm{vN}[\rho_{BH\cup R}]$.

Before starting the calculations, we note that the pure state \eqref{eq:puriSta} and the mixed state \eqref{eq:mixedstate} are related by tracing out the baby universe degrees of freedom
\begin{equation}
	\begin{aligned}
		\rho_{BH\cup R}&=\tr_{BU}\left[|\Phi\rangle \langle \Phi|_{BH\cup R\cup BU}\right]\\
		&=\sum_{M} p_M |\Psi_M\rangle \langle \Psi_M|_{BH\cup R}.
	\end{aligned}\label{eq:PureToMix}
\end{equation}
For accuracy, we explicitly give the probability distribution $p_M$ by (e.g., \cite{Kudler-Flam:2021alo,Kudler-Flam:2021rpr})
\begin{equation}
	\begin{aligned}
	p_M&= \left(\frac{\mathcal{N}k }{\pi}\right)^{\mathcal{N}k} \exp\left( - \mathcal{N}k \tr (C^M C^{M\dagger}) \right), 
	\end{aligned}\label{eq:probDis}
\end{equation}
and this probability distribution is normalized 
\begin{equation}
	\begin{aligned}
		\sum_M p_M  \to & \left(\frac{\mathcal{N}k }{\pi}\right)^{\mathcal{N}k}  \int \prod_{i,j=1}^{\mathcal{N}}\prod_{\alpha,\beta=1}^{k}d C^{M}_{i\alpha}d C^{\dagger M}_{\beta j} \; \exp\left( - \mathcal{N}k \tr (C^M C^{M\dagger}) \right)	\\
		&=1
	\end{aligned}\label{eq:normal}
\end{equation}
and gives the relationship \eqref{eq:haarRan}. Although we explicitly give the probability distribution, in calculating the entropies below, we do not use the explicit form \eqref{eq:probDis}, but the relationship \eqref{eq:haarRan}.

\subsection{The entropy of the naive Hawking radiation \texorpdfstring{$S_\mathrm{vN}[\rho_R]=S_\mathrm{vN}[\rho_{BH\cup BU}]$}{}} \label{subsec:Hawki}
To get the von Neumann entropy of the naive Hawking radiation $R$, we consider the reduced density matrix for the naive Hawking radiation.
It is given by 
\begin{equation}
	\begin{aligned}
			\rho_R&=\sum_{M} p_M\tr_{BH}\left[|\Psi_M\rangle \langle \Psi_M|_{BH\cup R} \right]\\
			&=\sum_M p_M \,\rho_{(M) R}\\
			&\equiv \langle\rho_{(M)R}\rangle_M,
	\end{aligned}
\end{equation}
where in the second line we defined the reduced density matrix 
\begin{equation}
	\begin{aligned}
	\rho_{(M)R}&=\tr_{BH}\left[|\Psi_M\rangle \langle \Psi_M|_{BH\cup R} \right]\\
	&=\sum_{i=1}^{\mathcal{N}}\sum_{\alpha,\beta=1}^{k}C^{M}_{i\alpha}C^{\dagger M}_{\beta i} | \alpha \rangle\langle \beta|_{R} \label{eq:ReducedR}
	\end{aligned}
\end{equation}
 and in the last line to emphasize the ensemble average of the reduced density matrix $\rho_{(M)R}$ we introduced the notation $\langle\rho_{(M)R}\rangle_M$ defined by the second line. We note that the average operation is given by the relationship \eqref{eq:haarRan} explicitly.

Next we consider the R\'{e}nyi entropy for the reduced density matrix
\begin{equation}
\begin{aligned}
		\tr_{R}\rho_{R}^{n}&=\sum_{M_1,M_2,\cdots,M_n}p_{M_1}p_{M_2}\cdots p_{M_n} \tr_{R}[\rho_{(M_1)R}\;\rho_{(M_2)R}\cdots \rho_{(M_n)R}]\\
		&=\sum_{M_1,M_2,\cdots,M_n}p_{M_1}p_{M_2}\cdots p_{M_n} \sum_{i_1,i_2,\cdots,i_n =1}^{\mathcal{N}} \sum_{\alpha_1,\alpha_2,\cdots,\alpha_n=1}^{k}C^{M_1}_{i_1 \alpha_1} C^{\dagger M_1}_{\alpha_2 i_1}C^{M_2}_{i_2 \alpha_2} C^{\dagger M_2}_{\alpha_3 i_2}\cdots C^{M_n}_{i_n \alpha_n} C^{\dagger M_n}_{\alpha_1 i_n}\\
		&=\sum_{i_1,i_2,\cdots,i_n =1}^{\mathcal{N}} \sum_{\alpha_1,\alpha_2,\cdots,\alpha_n=1}^{k} \la C^{M_1}_{i_1 \alpha_1} C^{\dagger M_1}_{\alpha_2 i_1}\ra_{M_1} \la C^{M_2}_{i_2 \alpha_2} C^{\dagger M_2}_{\alpha_3 i_2}\ra_{M_2} \cdots \la C^{M_n}_{i_n \alpha_n} C^{\dagger M_n}_{\alpha_1 i_n}\ra \\
		&=\sum_{i_1,i_2,\cdots,i_n =1}^{\mathcal{N}} \sum_{\alpha_1,\alpha_2,\cdots,\alpha_n=1}^{k} \frac{1}{(k\mathcal{N})^n} \delta_{i_1 i_1}\delta_{\alpha_1\alpha_2}\delta_{i_2 i_2}\delta_{\alpha_2\alpha_3}\cdots \delta_{i_n i_n}\delta_{\alpha_n\alpha_1}\\
		&=\frac{1}{k^{n-1}},
\end{aligned}
\end{equation}
where in the third line we distinguished the labels $M_1,\cdots,M_n$, take the ensemble averages for factors, which have the same label $M_s$, and in the forth line we used the relationship \eqref{eq:haarRan}. 

Therefore we obtain the von Neumann entropy \eqref{eq:vNcoase}
\begin{equation}
	\begin{aligned}
		S_\mathrm{vN}[\rho_R]&=S_\mathrm{vN}[\langle\rho_{(M)R}\rangle_M]\\
		&=-\lim_{n\to 1}\partial_n \tr_{R}\rho_{R}^{n}\\
		&=\log k.
	\end{aligned}
\end{equation}
This von Neumann entropy is coincides with the Hawking's result, and it implies the information paradox. We note that this von Neumann entropy is equal to that of the the union of the black hole and the baby universe $BH \cup BU$
\begin{equation}
	\begin{aligned}
		S_\mathrm{vN}[\rho_{BH \cup BU}]&=S_\mathrm{vN}[\rho_R]\\
		&=\log k.
	\end{aligned}
	\label{eq:SEERadiation}
\end{equation}

\subsection{The entropy of the naive Hawking radiation and the baby universe \texorpdfstring{\\ $S_\mathrm{vN}[\rho_{R\cup BU}]=S_\mathrm{vN}[\rho_{BH}]$}{}}\label{subsec:HawkiBaby}
To get the von Neumann entropy of the naive Hawking radiation and the baby universe $R \cup BU (=\mathbf{R})$, we consider the following  reduced density matrix
\begin{equation}
	\begin{aligned}
		\rho_{R\cup BU} &= \tr_{BH} \left[ |\Phi\rangle \langle \Phi|_{BH\cup R\cup BU} \right]\\
		&=\sum_{MN}\sqrt{p_M p_N} \left(\tr_{BH} |\Psi_M \ra \la \Psi_N |_{BH\cup R}\right) \otimes |M \ra \la N |_{BU}\\
		&= \sum_{MN}\sqrt{p_M p_N} \sum_{i=1}^{\mathcal{N}}\sum_{\alpha,\beta=1}^{k}C^{M}_{i\alpha}C^{\dagger N}_{\beta i} | \alpha \rangle\langle \beta|_{R}\otimes | M \rangle\langle N|_{BU}.
	\end{aligned}
\end{equation}

As in the previous case, we consider the R\'{e}nyi entropy
\begin{equation}
	\begin{aligned}
		\tr_{R\cup BU} \rho_{R\cup BU}^n &=\sum_{M_1,M_2,\cdots,M_n}p_{M_1}p_{M_2}\cdots p_{M_n}  \\
		& \hspace{2cm} \times\tr_{R}\left[ (\tr_{BH}|\Psi_{M_1}\ra \la \Psi_{M_2}|) (\tr_{BH}|\Psi_{M_2}\ra \la \Psi_{M_3}|) \cdots (\tr_{BH}|\Psi_{M_n}\ra \la \Psi_{M_1}|) \right]\\
		&=\sum_{M_1,M_2,\cdots,M_n}p_{M_1}p_{M_2}\cdots p_{M_n} \sum_{i_1,i_2,\cdots,i_n =1}^{\mathcal{N}} \sum_{\alpha_1,\alpha_2,\cdots,\alpha_n=1}^{k} C^{M_1}_{i_1 \alpha_1} C^{\dagger M_2}_{\alpha_2 i_1}C^{M_2}_{i_2 \alpha_2} C^{\dagger M_3}_{\alpha_3 i_2}\cdots C^{M_n}_{i_n \alpha_n} C^{\dagger M_1}_{\alpha_1 i_n}\\
		&=\sum_{i_1,i_2,\cdots,i_n =1}^{\mathcal{N}} \sum_{\alpha_1,\alpha_2,\cdots,\alpha_n=1}^{k} \la C^{\dagger M_1}_{\alpha_1 i_n} C^{M_1}_{i_1 \alpha_1}\ra_{M_1} \la  C^{\dagger M_2}_{\alpha_2 i_1}C^{M_2}_{i_2 \alpha_2}\ra_{M_2} \la  C^{\dagger M_3}_{\alpha_3 i_2} C_{i_3 \alpha _3}^{M_3} \ra_{M_3}\cdots  \la  C^{\dagger M_n}_{\alpha_n i_{n-1}} C^{M_n}_{i_n \alpha_n} \ra_{M_n}\\
		&=\sum_{i_1,i_2,\cdots,i_n =1}^{\mathcal{N}} \sum_{\alpha_1,\alpha_2,\cdots,\alpha_n=1}^{k} \frac{1}{(k\mathcal{N})^n} \delta_{i_n i_1}\delta_{\alpha_1\alpha_1}\delta_{i_1 i_2}\delta_{\alpha_2\alpha_2}\cdots \delta_{i_{n-1} i_n}\delta_{\alpha_n\alpha_n}\\
		&=\frac{1}{\mathcal{N}^{n-1}},
	\end{aligned}
\end{equation}
where in the fourth line we used the relationship \eqref{eq:haarRan}, and we note that $\mathcal{N}=e^{S_{BH}}$.

From this R\'{e}nyi entropy, we get the von Neumann entropy of the union of the naive Hawking radiation and the baby universe \eqref{eq:vNfine}
\begin{equation}
	\begin{aligned}
		S_\mathrm{vN}[\rho_{R\cup BU}]&=-\lim_{n\to 1}\partial_n \tr_{R\cup BU}\rho_{R\cup BU}^{n}\\
		&=\log \mathcal{N}\\
		&=S_{BH}.
	\end{aligned}
\end{equation}
This von Neumann entropy is also equal to that of the black hole $BH$
\begin{equation}
	\begin{aligned}
		S_\mathrm{vN}[\rho_{BH}]&=S_\mathrm{vN}[\rho_{R\cup BU}]\\
		&= S_{BH}.
	\end{aligned}
		\label{eq:SEEBH}
\end{equation}

\subsection{The entropy of the baby universe \texorpdfstring{$S_\mathrm{vN}[\rho_{BU}]=S_\mathrm{vN}[\rho_{BH \cup R}]$}{}}\label{subsec:Baby}
To get the von Neumann entropy of the baby universe $BU$, we consider the  following reduced density matrix,
\begin{equation}
	\begin{aligned}
		\rho_{BU}&= \tr_{BH\cup R} \left[ |\Phi\rangle \langle \Phi|_{BH\cup R\cup BU} \right]\\
		&=\sum_{MN}\sqrt{p_M p_N} \left(\tr_{BH\cup R} |\Psi_M \ra \la \Psi_N |_{BH\cup R}\right) \otimes |M \ra \la N |_{BU}\\
		&= \sum_{MN}\sqrt{p_M p_N} \sum_{i=1}^{\mathcal{N}}\sum_{\alpha=1}^{k}C^{M}_{i\alpha}C^{\dagger N}_{\alpha i}  | M \rangle\langle N|_{BU}.
	\end{aligned}\label{eq:densiRBU}
\end{equation}
Then we get the R\'{e}nyi entropy 
\begin{equation}
	\begin{aligned}
		\tr_{BU} \rho_{BU}^n&=\sum_{M_1,M_2,\cdots,M_n}p_{M_1}p_{M_2}\cdots p_{M_n} \sum_{i_1,i_2,\cdots,i_n =1}^{\mathcal{N}} \sum_{\alpha_1,\alpha_2,\cdots,\alpha_n=1}^{k} C^{M_1}_{i_1 \alpha_1} C^{\dagger M_2}_{\alpha_1 i_1}C^{M_2}_{i_2 \alpha_2} C^{\dagger M_3}_{\alpha_2 i_2}\cdots C^{M_n}_{i_n \alpha_n} C^{\dagger M_1}_{\alpha_n i_n}\\
		&=\sum_{i_1,i_2,\cdots,i_n =1}^{\mathcal{N}} \sum_{\alpha_1,\alpha_2,\cdots,\alpha_n=1}^{k} \la C^{\dagger M_1}_{\alpha_n i_n} C^{M_1}_{i_1 \alpha_1}\ra_{M_1} \la  C^{\dagger M_2}_{\alpha_1 i_1}C^{M_2}_{i_2 \alpha_2}\ra_{M_2} \la  C^{\dagger M_3}_{\alpha_2 i_2} C_{i_3 \alpha _3}^{M_3} \ra_{M_3}\cdots  \la  C^{\dagger M_n}_{\alpha_{n-1} i_{n-1}} C^{M_n}_{i_n \alpha_n} \ra_{M_n}\\
		&=\sum_{i_1,i_2,\cdots,i_n =1}^{\mathcal{N}} \sum_{\alpha_1,\alpha_2,\cdots,\alpha_n=1}^{k} \frac{1}{(k\mathcal{N})^n} \delta_{i_n i_1}\delta_{\alpha_n\alpha_1}\delta_{i_1 i_2}\delta_{\alpha_1\alpha_2}\cdots \delta_{i_{n-1} i_n}\delta_{\alpha_{n-1}\alpha_n}\\
		&=\frac{1}{(k \mathcal{N})^{n-1}},
	\end{aligned}
\end{equation}
where in the third line we used the rule \eqref{eq:haarRan}.

From the R\'{e}nyi entropy, we get the von Neumann entropy of the baby universe 
\begin{equation}
	\begin{aligned}
		S_\mathrm{vN}[\rho_{BU}]&=-\lim_{n\to 1}\partial_n \tr_{BU} \rho_{BU}^n\\
		&=\log (k \mathcal{N})\\
		&= S_{BH}+ \log k.
	\end{aligned}
\end{equation}
It is equal to the entropy of the union of the black hole and the Hawking radiation $BH\cup R$
\begin{equation}
	\begin{aligned}
		S_\mathrm{vN}[\rho_{BH \cup R}]&=S_\mathrm{vN}[\rho_{BU}]\\
		&=  S_{BH}+ \log k.
		\label{eq:SEEBU}
	\end{aligned}
\end{equation}

\bibliographystyle{JHEP}
\bibliography{Island}

\end{document}